\documentclass[useAMS,usenatbib]{mn2e}
\pdfoutput=1
\usepackage[pdftex,pdfpagemode={UseOutlines},bookmarks,bookmarksopen,colorlinks,linkcolor={blue},citecolor={green},urlcolor={red}]{hyperref}
\usepackage{graphicx}
\usepackage{xspace}

\usepackage{times}

\newcommand{\hmpc}{$\,{\rm h}^{-1}$ Mpc\xspace}
\newcommand{\wprp}{$w_p(r_p)$\xspace}
\newcommand{\kms}{$\,{\rm km \cdot s^{-1}}$\xspace}

\title[The zCOSMOS-Bright survey: the clustering of galaxy morphological types
  since $z\simeq 1$]{The zCOSMOS-Bright survey: the clustering of early and
  late galaxy morphological types since $z\simeq 1$}

\author[S. de la Torre et al.]
 {S. de la Torre,$^{1,2,3}$\thanks{E-mail: sylvain.delatorre@brera.inaf.it}
  O. Le F\`evre,$^{1}$
  C. Porciani,$^{4}$
  L. Guzzo,$^{2}$
  B. Meneux,$^{5,6}$
  U. Abbas,$^{7}$
  \newauthor
  L. Tasca,$^{3}$
  C.M. Carollo,$^{8}$
  T. Contini,$^{9}$
  J.-P. Kneib,$^{1}$
  S.J. Lilly,$^{8}$
  V. Mainieri,$^{10}$
  A. Renzini,$^{11}$
  \newauthor
  M. Scodeggio,$^{3}$
  G. Zamorani,$^{12}$
  S. Bardelli,$^{12}$
  M. Bolzonella,$^{12}$
  A. Bongiorno,$^{5}$
  K. Caputi,$^{8}$
  \newauthor
  G. Coppa,$^{12}$
  O. Cucciati,$^{1}$
  L. de Ravel,$^{1}$
  P. Franzetti,$^{3}$
  B. Garilli,$^{3}$
  C. Halliday,$^{13}$
  A. Iovino,$^{2}$
  \newauthor
  P. Kampczyk,$^{8}$
  C. Knobel,$^{8}$
  A.M. Koekemoer,$^{14}$
  K. Kova{\v c},$^{8}$
  F. Lamareille,$^{9}$
  J.-F. Le Borgne,$^{9}$
  \newauthor  
  V. Le Brun,$^{1}$
  C. Maier,$^{8}$
  M. Mignoli,$^{12}$
  R. Pell\'o,$^{9}$
  Y. Peng,$^{8}$
  E. Perez-Montero,$^{9}$
  \newauthor
  E. Ricciardelli,$^{15}$
  J. Silverman,$^{16}$
  M. Tanaka,$^{10}$
  L. Tresse,$^{1}$
  D. Vergani,$^{12}$
  E. Zucca,$^{12}$
  \newauthor
  D. Bottini,$^{3}$
  A. Cappi,$^{12}$
  P. Cassata,$^{17}$
  A. Cimatti,$^{18}$
  A. Leauthaud,$^{19}$
  D. Maccagni,$^{3}$
  \newauthor
  C. Marinoni,$^{20}$
  H.J. McCracken,$^{21}$
  P. Memeo,$^{3}$
  P. Oesch,$^{8}$
  L. Pozzetti$^{12}$
  and R. Scaramella$^{22}$ \\
  $^{1}$ Laboratoire d'Astrophysique de Marseille, 13388 Marseille, France \\
  $^{2}$ INAF - Osservatorio Astronomico di Brera, 23807 Merate, Italy \\
  $^{3}$ INAF - Istituto di Astrofisica Spaziale e Fisica Cosmica di Milano, 20133 Milano, Italy \\
  $^{4}$ Argelander Institute for Astronomy, University of Bonn, 53121 Bonn, Germany \\
  $^{5}$ Max Planck Institut f\"ur Extraterrestrische Physik, 85748 Garching, Germany \\
  $^{6}$ Universitats-Sternwarte, 81679 Munich, Germany \\
  $^{7}$ INAF - Osservatorio Astronomico di Torino, 10025 Pino Torinese, Italy \\
  $^{8}$ Institute of Astronomy, ETH Zurich, 8093 Zurich, Switzerland \\
  $^{9}$ Laboratoire d'Astrophysique de l'Observatoire Midi-Pyr\'en\'ees, 31400 Toulouse, France \\
  $^{10}$ European Southern Observatory, 85748 Garching, Germany \\
  $^{11}$ INAF - Osservatorio Astronomico di Padova, 35122 Padova, Italy \\
  $^{12}$ INAF - Osservatorio Astronomico di Bologna, 40127 Bologna, Italy \\
  $^{13}$ INAF - Osservatorio Astrofisico di Arcetri, 50125 Firenze, Italy \\
  $^{14}$ Space Telescope Science Institute, 21218 Baltimore, USA \\
  $^{15}$ Dipartimento di Astronomia, Universit\'a di Padova, 35122 Padova, Italy \\
  $^{16}$ Institute for the Physics and Mathematics of the Universe, University of Tokyo, Kashiwa-shi, Chiba 277-8568, Japan \\  
  $^{17}$ Department of Astronomy, University of Massachusetts, 01003 Amherst, USA \\
  $^{18}$ Dipartimento di Astronomia, Universit\'a di Bologna, 40127 Bologna, Italy \\
  $^{19}$ Berkeley Lab \& Berkeley Center for Cosmological Physics, University of California, 94720 Berkeley, USA \\
  $^{20}$ Centre de Physique Th\'eorique de Marseille, 13288 Marseille, France \\
  $^{21}$ Institut d'Astrophysique de Paris, 75014 Paris, France \\
  $^{22}$ INAF - Osservatorio Astronomico di Roma, 00040 Monte Porzio Catone, Italy}

\begin{document}

\date{Accepted 2010 October 28. Received 2010 October 23; in original form 2009 October 23}

\pagerange{\pageref{firstpage}--\pageref{lastpage}} \pubyear{2010}

\maketitle

\label{firstpage}

\begin{abstract}

We measure the spatial clustering of galaxies as a function of their
morphological type at $z\simeq0.8$, for the first time in a deep redshift
survey with full morphological information. This is obtained by combining
high-resolution HST imaging and VLT spectroscopy for about $8,500$ galaxies to
$I_{AB}=22.5$ with accurate spectroscopic redshifts from the zCOSMOS-Bright
redshift survey. At this epoch, early-type galaxies already show a
significantly stronger clustering than late-type galaxies on all probed
scales. A comparison to the SDSS at $z\simeq0.1$, shows that the relative
clustering strength between early and late morphological classes tends to
increase with cosmic time at small separations, while on large scales it shows
no significant evolution since $z\simeq0.8$. This suggests that most
early-type galaxies had already formed in intermediate and dense environments
at this epoch. Our results are consistent with a picture in which the relative
clustering of different morphological types between $z\simeq1$ and $z\simeq0$,
reflects the evolving role of environment in the morphological transformation
of galaxies, on top of the global mass-driven evolution.

\end{abstract}


\section{Introduction}

In the local Universe, the clustering properties of galaxies depend on
luminosity
\citep[e.g.][]{davis88,hamilton88,white88,park94,loveday95,benoist96,
  guzzo00,norberg01,zehavi02,zehavi05,li06,skibba06,wang07,swanson08}, colour
\citep[e.g.][]{willmer98,zehavi02,zehavi05,li06,swanson08,skibba09c}, spectral
type \citep[e.g.][]{loveday99,norberg02,madgwick03,wang07}, and environment
\citep{abbas06,abbas07}. Differences in the clustering of the various galaxy
morphological types are also observed
\citep[e.g.][]{davis76,giovanelli86,iovino93,loveday95,hermit96,guzzo97,
  willmer98,zehavi02,skibba09}. Elliptical galaxies are more strongly
clustered than spiral and irregular galaxies. Another manifestation of this
phenomenon is the existence of a morphology-density relation
\citep{dressler80,postman84}, which implies that a higher fraction of
ellipticals than either spirals or irregulars, reside in denser
environments. Spiral and irregular galaxies are more likely to populate less
dense regions. However, the origin and relation between these dependences are
still not fully understood. They are usually discussed in terms of a
\emph{nature} or \emph{nurture} scenario, i.e. being intrinsic properties of
galaxies at formation or originated from the interaction of galaxies with
their environment all along their cosmic evolution.

At higher redshifts, our knowledge of galaxy clustering properties remains
fragmentary. We know little about the clustering of the various morphological
types. Deep surveys have detected an evolution in the global galaxy clustering
with redshift up to $z\sim1-1.5$
\citep{coil04,lefevre05,mccracken08}. Additional results indicate that the
luminosity, colour, and spectral type dependences of clustering evolve with
redshift. A less significant segregation is evident when we consider the
dependence of clustering on either colour or spectral type, the difference in
clustering strength between red/early-type and blue/late-type galaxies being
shallower at $z\sim1-1.5$ than at $z\simeq0$ \citep{meneux06,coil08}.

These observations are consistent with the accepted picture of hierarchical
clustering growth and galaxy evolution, in which cold dark matter haloes form
from the gravitational collapse of dark matter around peaks in the density
field. Haloes evolve hierarchically, such that smaller haloes assemble to form
larger and more massive haloes in high-density regions
\citep{mo96,sheth02}. In parallel, galaxies form within haloes, by means of
the cooling of hot baryonic gas \citep{white87}. In this framework, luminous
and massive galaxies are expected to be more strongly clustered than fainter
and less massive ones that tend to form in less clustered haloes, which are
less biased with respect to the underlying mass distribution. Moreover,
early-type galaxies, which are in general brighter than late-type galaxies,
are more strongly clustered. This difference is thought to be related to their
different formation histories. Early-type galaxies are understood to be the
product of interaction processes (e.g. galaxy merging) and/or physical
mechanisms that take place in dense environments, although mergers are more
common in less dense group environments than at galaxy cluster cores
\citep{ellison10}. Instead, late-type galaxies may not experience such
dramatic events, being more likely to reside in less dense environments.
Mergers are expected to affect galaxy morphology by creating or growing a
galaxy bulge component, albeit a large gas fraction may result in a galaxy
temporarily re-growing a disk. They contribute to the evolving number
densities \citep[e.g.][]{deravel09} and clustering properties of different
types of galaxies.

Galaxy colours or spectral types are often used as proxies for galaxy
morphologies, but are affected by uncertainties in tracing the same underlying
galaxy populations across cosmic time. For instance, a disk galaxy containing
an old stellar population would be classified as an early-type object. In a
similar way, a starburst galaxy heavily obscured by dust would be classified
as a red, early-type galaxy. In contrast, an old and red stellar population
may dominate the measured stellar mass of an object, but the galaxy colour can
be blue because of a recent burst of star formation. Colours are primarily
related to galaxy recent star formation history at variance with morphologies,
which may be the result of the interaction of galaxies with their
surroundings. Selecting galaxies by colour or spectral type, or by morphology
may give one different views on galaxy evolution and environmental effects at
work. Studying how galaxy clustering depends on morphology and cosmic epoch
may thus provide important and complementary clues to mechanisms through which
galaxies developed into the population we see today. Such a study has not yet
been possible at $z>0.3$ because of the lack of high resolution imaging
surveys of sufficient size. The Cosmic Evolution Survey
\citep[COSMOS,][]{scoville07} provides us with a unique sample in this
respect.

In this work, we use the zCOSMOS-Bright spectroscopic sample of galaxies
\citep{lilly09} and measure for the first time the morphological dependence of
galaxy clustering at $z\simeq0.8$.  This paper is part of a series of three
using the first epoch zCOSMOS-Bright sample, that investigate galaxy
clustering as a function of galaxy physical properties: morphology (this
paper), luminosity and stellar mass \citep{meneux09}, and colour (Porciani et
al., in preparation).

The paper is organised as follows. In Sect. 2, we present the galaxy sample
and its basic properties. In Sect. 3, we describe the method used to measure
galaxy clustering. In Sect. 4, we provide the measurements of early- and
late-type galaxy clustering. In Sect. 5, we summarise and discuss our
results. Throughout this paper, we assume a flat $\Lambda CDM$ cosmology with
$\Omega_M=0.25$, $\Omega_\Lambda=0.75$, and $H_0=100~\rm{h~km \cdot s^{-1}
  \cdot Mpc^{-1}}$. The magnitudes are quoted in the AB system and for
simplicity, we denote the absolute magnitude in $B$-band $M_B-5\log{\rm (h)}$
as $M_B$.

\section{The data}

\subsection{The zCOSMOS-Bright catalogue}

zCOSMOS is an ongoing large spectroscopic survey being performed with the
Visible Multi-Object Spectrograph \citep[VIMOS,][]{lefevre03} at the European
Southern Observatory's Very Large Telescope (ESO-VLT). The bright part of the
survey, zCOSMOS-Bright, has been designed to follow-up spectroscopically the
entire $1.7~\rm{deg}^2$ COSMOS-ACS field \citep{scoville07,koekemoer07} to as
faint as $I_{AB}=22.5$. Observations have used the medium resolution red grism
with $1$ arcsec slits, yielding a spectral resolution of ${\rm R}\sim600$ at
2.5 {\rm \AA}~${\rm pixel}^{-1}$. The velocity uncertainty in the redshifts is
estimated to be $110$\kms. The survey strategy, which consists of 8 passes of
the VIMOS spectrograph across the field, will allow us to reach in the end a
high sampling rate of about $70\%$ over the entire field \citep{lilly07}. In
the present analysis, we use the first epoch zCOSMOS-Bright spectroscopic
sample (also called \emph{zCOSMOS 10k-bright sample}), which is based on the
first two years of observations performed in 2005-2006. The first epoch sample
covers an effective area of $1.5~\rm{deg}^2$ with an average sampling rate of
about $30\%$ across the field. It consists of $10,644$ magnitude-selected
objects according to $I_{AB}<22.5$.  This selection provides a redshift
distribution in the range $0.1<z<1.2$ that peaks at $z\simeq0.6$. A confidence
class has been assigned to each object of the sample to characterise the
confidence in the redshift determination and its nature. In this work, we use
only galaxies with secure redshifts, i.e., objects with confidence classes
$4.\mathsf{x}$, $3.\mathsf{x}$, $9.3$, $9.5$, $2.4$, $2.5$, and $1.5$. Our
sample thus represents $88\%$ of the full first-epoch zCOSMOS-Bright catalogue
\citep[see][for details]{lilly09}. The zCOSMOS survey is able to benefit from
the unprecedented multiwavelength coverage of the COSMOS field
\citep{capak07}, which permits us to compute very accurate photometric
redshifts \citep[e.g.][]{ilbert09}. By incorporating the photometric redshift
consistency information, we reach a spectroscopic confirmation rate of
$99\%$. The zCOSMOS survey design and the basic properties of the
zCOSMOS-Bright catalogue are fully described in \citet{lilly07} and
\citet{lilly09}.

\begin{figure}
\resizebox{\hsize}{!}{\includegraphics{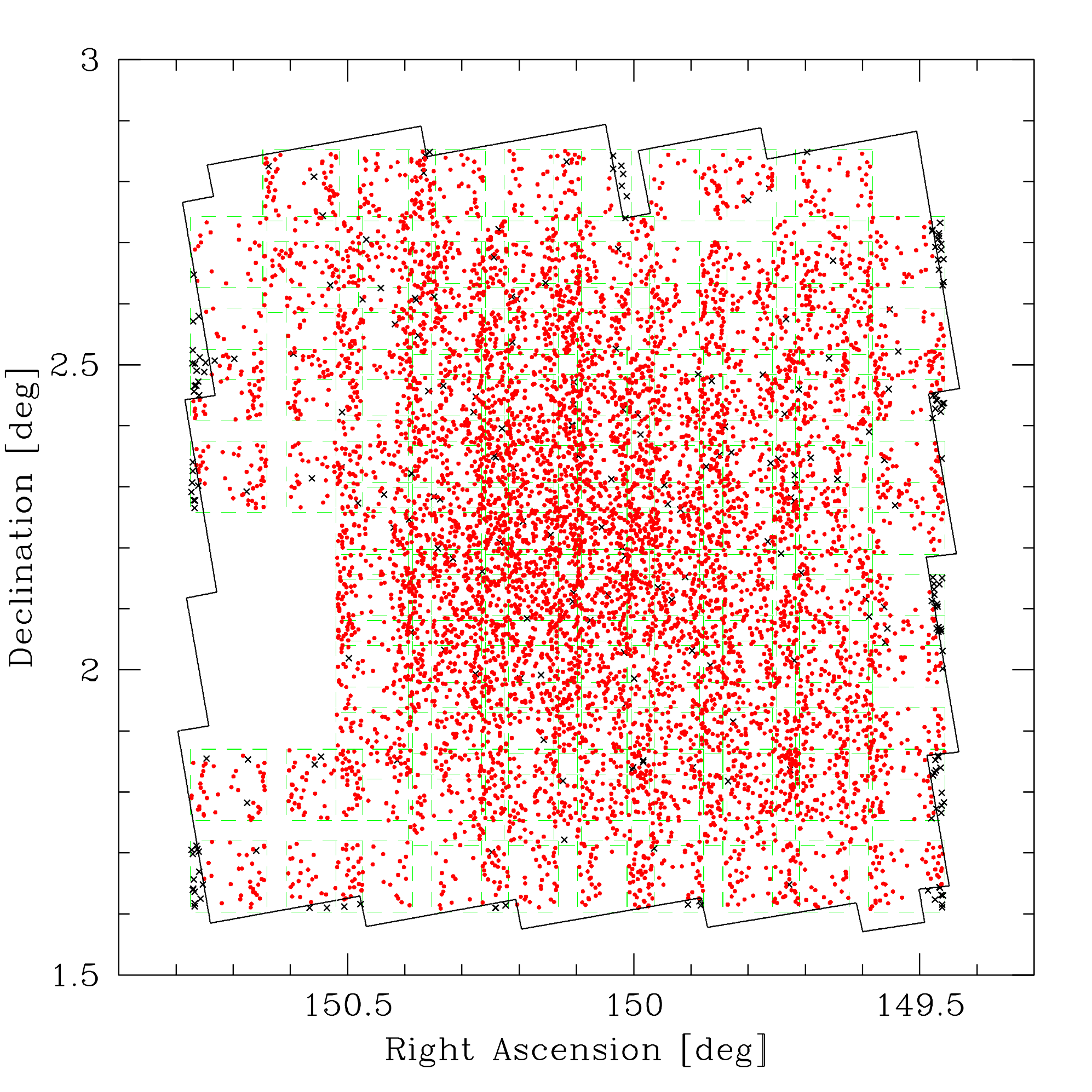}}
\caption{Angular distribution of the zCOSMOS-Bright sample. The solid line
  shows the boundary of the COSMOS-ACS survey and the dashed contours encompass
  the regions observed by VIMOS during the zCOSMOS-Bright first epoch
  observations. Crosses indicate the few galaxies for which no morphological
  information is available because either they lie outside the COSMOS-ACS
  field or coincide with photometric defects in the images.}
\label{sample}
\end{figure}

\subsection{Galaxy morphology}

A unique advantage of the COSMOS/zCOSMOS surveys is the HST-ACS high
resolution imaging \citep{koekemoer07} that is available over an area of
$1.7~\rm{deg}^2$. In particular, these data allow us to determine directly the
morphology of galaxies out to $z>1$. Morphologies used in this work were
estimated using a robust classification scheme based on three non-parametric
diagnostics of galaxy structure: concentration, asymmetry, and the Gini
coefficient \citep[][]{abraham96,abraham03,lotz04}. We first computed the
distance of each galaxy in the multi-parameter space with respect to each
galaxy of a control sample of 500 galaxies that had been visually
classified. A morphological class was then assigned to each galaxy depending
on the most frequent class among those of the $11^{th}$ nearest neighbours in
the control sample. The use of the $11^{th}$ nearest neighbours has been
chosen as it maximises the completeness of the different classes in our
sample. Galaxies were classified into three main morphological types:
early-type, spiral, and irregular, with the early-type including both
elliptical and lenticular galaxies (S0).

The classifier was optimised to be insensitive to band-shifting and surface
brightness dimming. Since the classification is performed using a single
photometric band and we study a relatively broad range of redshifts, our
morphology measurements may be biased as a function of redshift because of the
band-shifting effect \citep[e.g.][]{lotz04}. However, as discussed in
\citet{tasca09} our morphologies can only be affected at redshifts greater
than $z\simeq0.8$. Furthermore, the effect is small when only two broad
classes of early- and late-type galaxies are defined \citep{brinchmann98},
which is the case here. In practice, spiral galaxies can be misclassified as
irregulars but, since we define a broad class of late-type galaxies that
includes both spiral and irregular galaxies, our analysis is unaffected by
this bias. We evaluated the contamination and completeness of the different
classes by applying our automated classifier to the control sample and
comparing the resulting classification with the visual one. This comparison
showed that our classification scheme is rather robust with only a small
contamination of about $7\%$ for each class when considering only two broad
classes of early-type and spiral+irregular galaxies. We verified, using more
conservative morphological classes with almost no contamination but larger
incompleteness \citep[i.e. the ``clean'' sample of][]{tasca09}, that this has
a limited impact on the clustering measurements. The effect is much smaller
than the statistical error on \wprp measurements. We refer the reader to
\citet{tasca09} for a full description of the classification scheme and the
properties of the morphological sample.

In this study we focus on two broad classes: early types and late types, the
latter referring to galaxies morphologically classified as either spirals or
irregulars. Because of the slightly different geometries of the COSMOS-ACS and
zCOSMOS-Bright fields, a small number of galaxies in the zCOSMOS-Bright
catalogue do not have morphological measurements. These galaxies, which
represent $3\%$ of the zCOSMOS-Bright sample, are indicated by crosses in
Fig. \ref{sample}. In this analysis, we therefore remove galaxies that do not
have a counterpart in the COSMOS-ACS photometric catalogue
\citep{leauthaud07}. In the end, the effective area of the sample that we can
use is defined to be the intersection of the COSMOS-ACS and zCOSMOS fields,
after removal of regions affected by photometric defects for which no
morphological measurement is available (see Fig. \ref{sample}).

\subsection{Sample selection}

We consider a nearly volume-limited sample where galaxies are selected to be
of absolute magnitude below a given threshold and to populate a complete
comoving volume in the considered redshift interval. We focus on the
high-redshift part of the zCOSMOS-Bright sample, defining volume-limited
samples of early- and late-type galaxies at $0.6<z<1.0$. The selection is
performed on the basis of the galaxy $B$-band absolute magnitudes to
facilitate comparison with other studies. We include an intrinsic luminosity
evolution in the definition of the sample by considering a linearly evolving
absolute magnitude cut such as
\begin{equation}
M_B<-20.4-Q(1.0-z),
\end{equation}
where $Q$ is a constant. This is performed to ensure that the evolution in the
characteristic galaxy luminosity within the considered redshift interval is
taken into account. This absolute magnitude threshold allows us to select
galaxies with luminosities above $L^*$ at these redshifts. The mean luminosity
of all galaxies selected above this threshold is about $1.1L^*$ \citep{zucca09}.

\begin{figure}
\includegraphics[width=84mm]{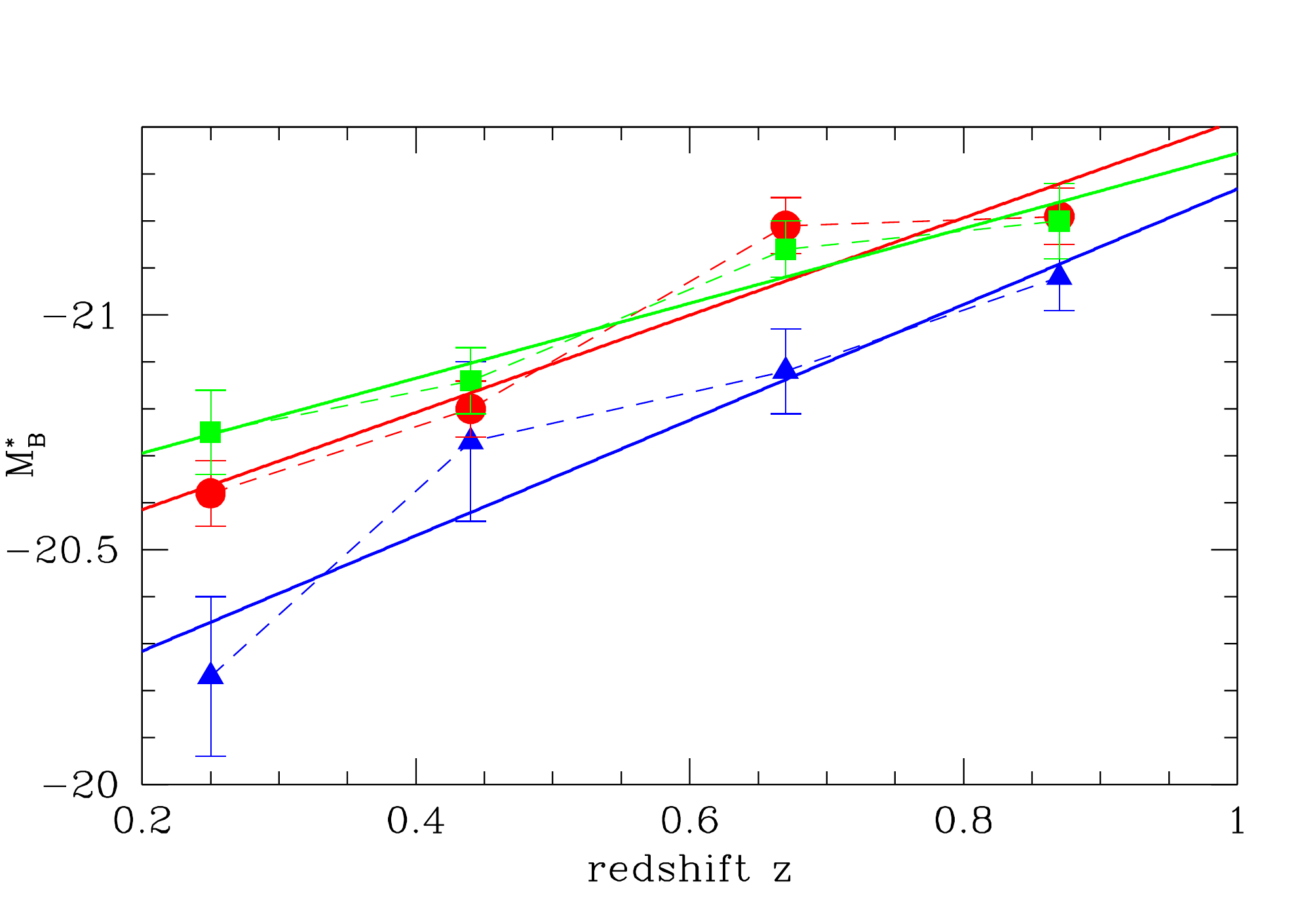}
\caption{Evolution of the characteristic absolute magnitude in B-band,
  $M_B^*$, as a function of redshift as measured by \citet{zucca09} for
  early-type (circles), spiral (squares), and irregular (triangles) galaxies
  in the zCOSMOS-Bright sample. The lines correspond to the best-fitted linear
  evolution of $M_B^*(z)$ for the three morphological types.}
\label{mstar_evol}
\end{figure}

Different galaxy populations exhibit different luminosity evolutions
\citep[][]{lilly96,zucca06} and thus $Q$ depends on the morphological type.
The value of $Q$ can be inferred observationally from the evolution of the
luminosity function for the different morphological types. We used the
characteristic absolute magnitude in the $B$-band, $M_B^*$, derived from the
same data by \citet{zucca09} in different redshift intervals and for different
morphological types. We note that we used $M^*$ values obtained by keeping the
faint-end slope of the luminosity function $\alpha$ fixed. This is a
reasonable decision since $\alpha$ does not clearly exhibit any type of
evolution at these redshifts \citep{zucca09}. We adopted this approach to
avoid compensating for the luminosity evolution with an artificial increase in
$\alpha$.

We therefore determined $Q$ by fitting the redshift evolution in $M_B^*$ with
a linear function. The values of $M_B^*(z)$ for the three morphological types
and their best-fit linear functions are presented in Fig. \ref{mstar_evol}. We
find a small amount of luminosity evolution for early types and spirals,
given, respectively, by $Q_E=-1.04\pm0.22$ and $Q_S=-0.80\pm0.14$, but a
stronger evolution for irregular galaxies, for which $Q_I=-1.23\pm0.24$. In
the case of irregular galaxies, the value of $Q$ is driven significantly in
the fit by the low-redshift point, which has a relatively small and unexpected
value. Excluding this point in the fit yields to a higher value of
$Q_I=-0.84\pm0.11$, close to that found for spirals. Therefore, after several
tests and given the large uncertainty in these values, we assumed that $Q=-1$
for all morphological types, which is close to the average of the three
morphological classes. We directly tested that varying the slope of the
absolute magnitude cut removes or adds only a small fraction of galaxies and
does not influence significantly the measurements of the correlation function.

\begin{figure}
\includegraphics[width=84mm]{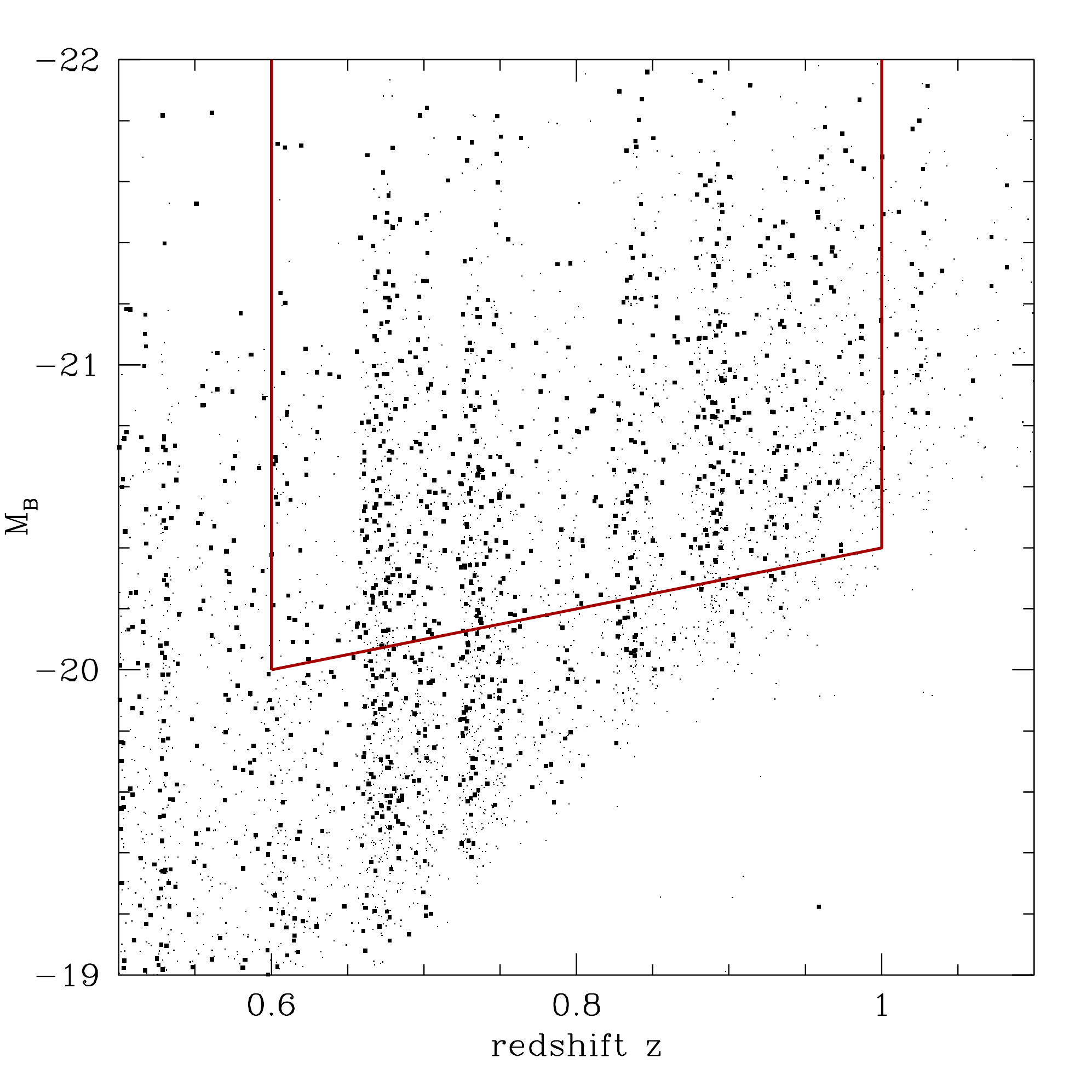}
\caption{Distribution of the zCOSMOS-Bright galaxies in the $B$-band absolute
  magnitude--redshift plane. Early-type galaxies are indicated by squares and
  late-type galaxies by dots. The solid line shows the boundary of the defined
  volume-limited sample.}
\label{Mze}
\end{figure}

\section{Galaxy clustering estimation}

\subsection{The projected two-point correlation function} \label{method}

To estimate galaxy clustering, we use the standard projected two-point
projected correlation function $w_p(r_p)$, which enables us to measure
real-space clustering without being affected by redshift-space
distortions. The three-dimensional galaxy distribution that is recovered from
redshift surveys and its two-point correlation function $\xi(s)$ are distorted
because of the effect of galaxy peculiar motions. These motions affect in
particular the distance measurements in the radial direction. By dividing the
vector $s$ into two components, $r_p$ and $\pi$, respectively perpendicular
and parallel to the line-of-sight \citep{peebles80,fisher94}, one obtains the
bi-dimensional two-point correlation function $\xi(r_p,\pi)$. One can then
recover the true spatial correlations by integrating over the redshift-space
distortion field, projecting $\xi(r_p,\pi)$ along the line-of-sight. The
projected two-point correlation function $w_p(r_p)$ is related to both
$\xi(r_p,\pi)$ and the real-space two-point correlation function $\xi(r)$ by
the relations
\begin{equation}
  w_p(r_p)=2\int_{0}^{\infty}\xi(r_p,\pi)d\pi
  =2\int_{r_p}^{\infty}\frac{\xi(r)}{\left(r^2-r_p^2\right)^{1/2}}rdr . \label{WPRP}
\end{equation}
To compute $w_p(r_p)$, we chose in practice to integrate $\xi(r_p,\pi)$ out to
$\pi_{max}=20$\hmpc. Given the volume of the survey, we found that this value
is sufficiently large and optimally minimises the noise introduced at large
$\pi$ by the uncorrelated pairs in the data \citep[see also][Porciani et al.,
  in preparation]{meneux09}. This truncation introduces a global
underestimation of $w_p(r_p)$ of about $10\%$ that has to be taken into
account when modelling the observations.

We computed $\xi(r_p,\pi)$ using the standard \citet{landy93} estimator, which
is defined by
\begin{equation}
  \xi(r_p,\pi)=\frac{GG(r_p,\pi)-2GR(r_p,\pi)+RR(r_p,\pi)}{RR(r_p,\pi)}, \label{LS} 
\end{equation}
where $GG(r_p,\pi)$, $RR(r_p,\pi)$, and $GR(r_p,\pi)$ are, respectively, the
normalised numbers of distinct galaxy-galaxy, galaxy-random, and random-random
pairs with comoving separations between $[r,r + dr]$ and $[\pi,\pi+d\pi]$. In
this analysis, we used random samples of $90,000$ objects.

\subsection{Observational biases}

\begin{figure}
\resizebox{\hsize}{!}{\includegraphics{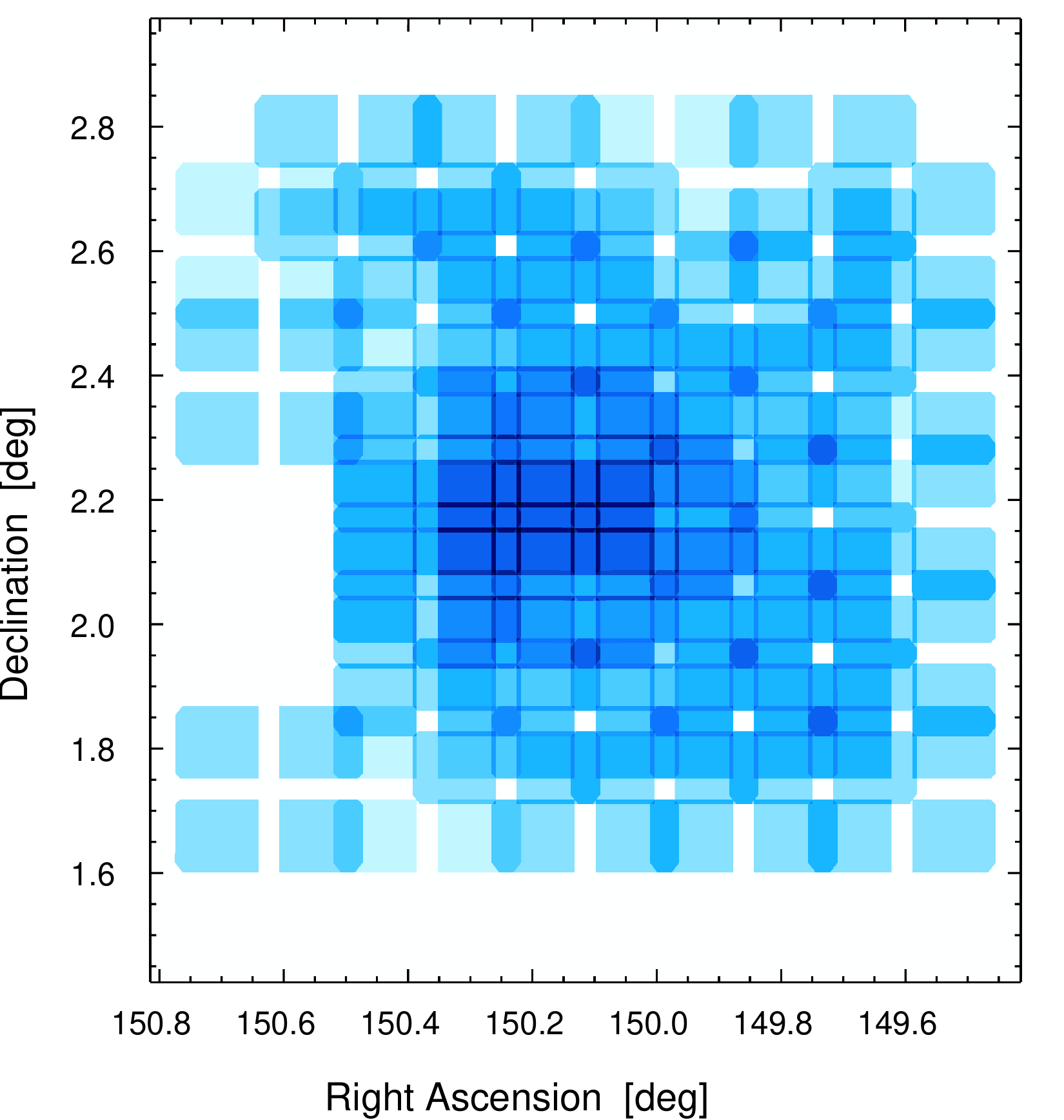}}
\caption{Variation in the angular sampling of the zCOSMOS-Bright sample. The
  colour darkness encodes the number of passes of the VIMOS spectrograph on
  the sky during zCOSMOS-Bright first epoch observations. The dark colour at
  the centre of field corresponds to the VIMOS 8-passes coverage expected over
  the full area in the final zCOSMOS-Bright sample.}
\label{zcsamp}
\end{figure}

The zCOSMOS-Bright sample has a complex angular sampling because of the survey
observational strategy and the shape of the VIMOS field-of-view. In
Fig. \ref{sample}, one can discern the VIMOS footprint and the superposition
of multiple passes in the angular distribution of galaxies. To correct the
projected correlation function for this non-uniform and incomplete angular
sampling, we accurately estimated the background counts expected for
unclustered objects in the field. We generated random catalogues with the
detailed angular selection function of the sample by varying the number of
random galaxies with angular position. More precisely, from the knowledge of
the precise shape of the VIMOS field-of-view and the coordinates of the
observed pointings, we distribute iteratively random galaxies inside the areas
covered by each pointing. With this procedure, we reproduce in the random
sample the effective surface density variations due to the non-uniform
spectroscopic sampling of the zCOSMOS sample, illustrated in
Fig. \ref{zcsamp}. The objects within regions of the parent photometric
catalogue affected by photometric defects, and for which no morphological
measurements were possible, were removed a posteriori by applying the
COSMOS-ACS photometric mask \citep{leauthaud07}. This technique was already
applied to the clustering analysis of the VIMOS-VLT Deep Survey (VVDS)
\citep[e.g.][]{lefevre05,pollo06,meneux06,delatorre07}, which suffers from
similar observational biases \citep[see][for details]{pollo05}. In the present
analysis, we improve this method by adding a correction for the non-uniform
sampling inside each pointing and including a more accurate pair-weighting
scheme to correct for the angular incompleteness, as described below.

\begin{figure}
\includegraphics[width=84mm]{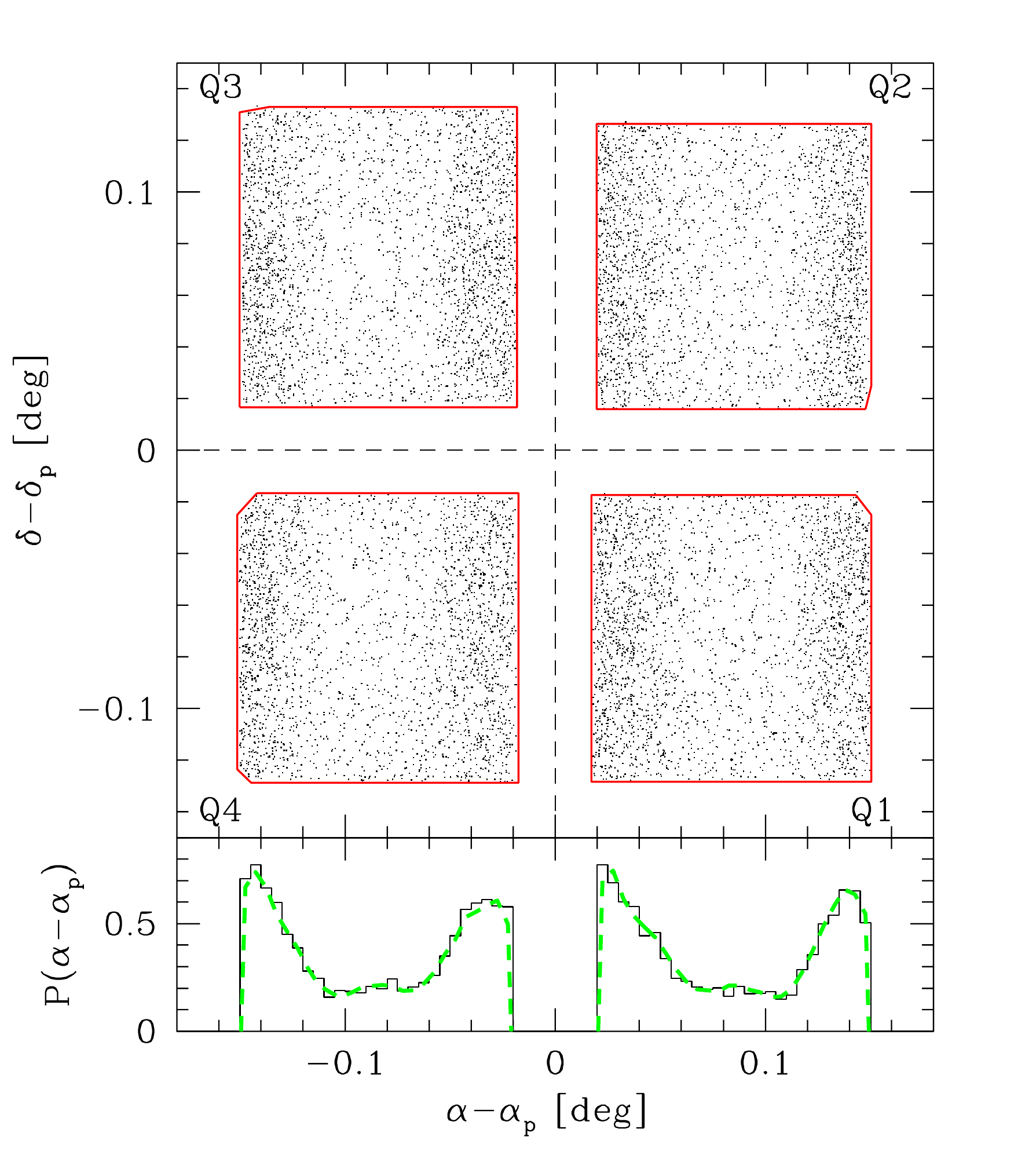}
\caption{Top panel: Distribution of the zCOSMOS-Bright galaxies in the
  rest-frame angular coordinates of their pointing, showing the uneven
  coverage of each of the four VIMOS quadrants. Bottom panel: corresponding
  histogram along the right ascension direction. The dashed curve is the
  smooth function used to model the average galaxy distribution along the
  right ascension direction within each pointing.}
\label{alprof}
\end{figure}

In the top panel of Fig. \ref{alprof}, we plot the distribution of
zCOSMOS-Bright galaxies in the rest-frame angular coordinates of their
pointing. The solid contours correspond to the effective shape of the four
quadrants of the VIMOS field-of-view. In this figure, one can discern a
variation in sampling as a function of the position within the pointing, in
the right ascension direction. This non-uniform sampling is produced during
the designing of slit masks to optimise the total number of slits, and in turn
produces two layers of spectra in each mask \citep{bottini05}. This problem is
more severe for zCOSMOS-Bright than VVDS because of the use of a higher
resolution grism and consequently longer spectra on the detector, which limit
the degree of freedom of the slit positioning algorithm. We therefore correct
for this by reproducing within the random samples the
right-ascension-dependent sampling of each pointing. We model the average
distribution of galaxies along the right ascension direction with a smooth
function, as shown in the bottom panel of Fig. \ref{alprof}, and use the
fitted model to distribute the galaxies in the random sample.

\begin{figure}
\includegraphics[width=84mm]{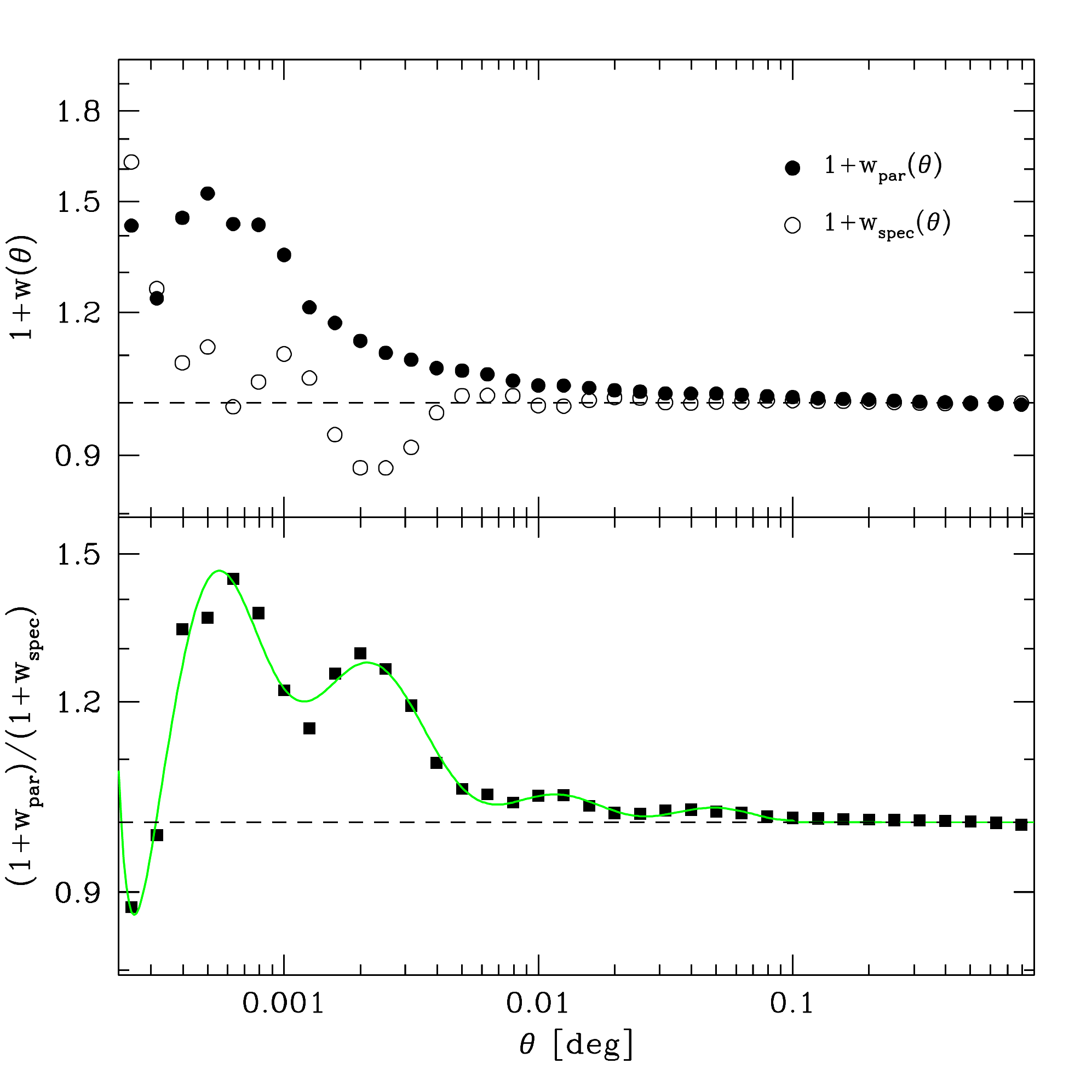}
\caption{Top panel: angular correlation functions of the parent photometric
  catalogue $w_p(\theta)$ (solid points) and the spectroscopic catalogue
  $w_s(\theta)$ (open points). Bottom panel: ratio of these two functions,
  where the weighting function (solid curve) was used to correct for the
  angular incompleteness.}
\label{corrweight}
\end{figure}

In addition, we use a pair-weighting scheme to account for the small-scale
incompleteness caused by the geometrical constraints of the slit mask design
\citep{bottini05}, which prevent VIMOS from simultaneously targeting very
close objects. As shown in the upper panel of Fig. \ref{corrweight}, in
general this incompleteness affects the angular clustering on scales of
$\theta<0.02~{\rm deg}$, but the effect becomes significant only below
$\theta=0.004~{\rm deg}$. The latter angular scale corresponds to a maximum
separation of $r_p=0.17$\hmpc at $0.6<z<1.0$. To correct for this, we assign a
weight to each galaxy-galaxy pair that accounts for the missed angular
pairs. To calculate the weights, we use the information contained in the
parent photometric catalogue, which is assumed to be free from angular
incompleteness. We define a weighting function $f(\theta)$ to be the ratio of
the mean number of pairs in the parent photometric catalogue to those in the
spectroscopic catalogue as a function of angular separation. This quantity can
be written as \citep[e.g.][]{hawkins03},
\begin{equation}
  f(\theta)=\frac{1+w_{par}(\theta)}{1+w_{spec}(\theta)}
\end{equation}   
where $w_{par}(\theta)$ and $w_{spec}(\theta)$ are the angular correlation
functions of the parent photometric and spectroscopic samples. We use again
the \citet{landy93} estimator to compute the angular correlation functions. In
the calculation of $w_{spec}(\theta)$, we account for the complex angular
sampling of the spectroscopic catalogue, as discussed above. We note that only
the galaxy-galaxy pairs are weighted because the random sample does not have
any close-pair constraints. The weighting function and the angular correlation
functions of the parent photometric and spectroscopic samples are shown in
Fig. \ref{corrweight}. We compared this weighting scheme with those adopted in
the parallel clustering analyses by \citet{meneux09} and Porciani et al. (in
preparation), and found that the different estimators yield consistent results
within the errors. More detailed results of this comparison are presented in
Porciani et al. (in preparation).

\begin{figure}
\includegraphics[width=84mm]{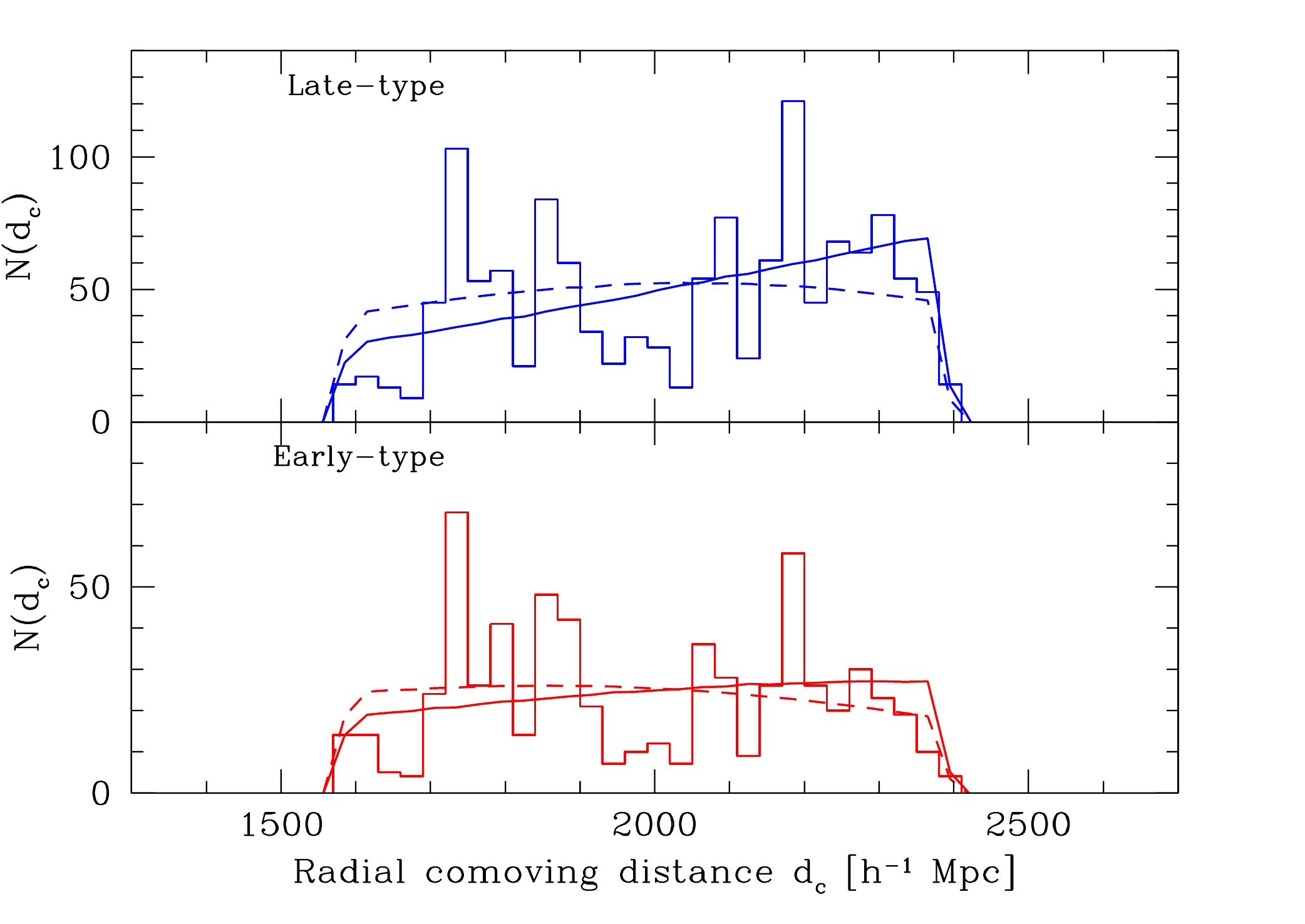}
\caption{Observed radial distribution in our volume-limited sample at
  $0.6<z<1.0$ (histogram) for late- (upper pannel) and early-type (lower
  panel) galaxies. The dashed curve shows the distribution obtained by
  smoothing the observed radial distribution with a Gaussian filter using a
  smoothing length of $\sigma_s=450$\hmpc. The solid curve corresponds
  to the distribution derived by integrating the galaxy luminosity function.}
\label{smoothing}
\end{figure}

Finally, to compute $\xi(r_p,\pi)$, the random catalogues need also to
reproduce the detailed radial selection function of the survey. Galaxies are
clustered and, because the volumes probed by galaxy surveys are usually small,
it is difficult to measure average quantities such as the radial distribution
of objects. The subsequent variability in the measurements, usually referred
to as sample variance (or cosmic variance), can significantly bias its
determination \citep[e.g.][]{garilli08}. One way of obtaining the radial
selection function entails convolving the observed galaxy redshift
distribution with a kernel of sufficiently large width to smooth out the
resolved structures. The zCOSMOS-Bright sample contains a large number of
structures along the line-of-sight that correspond to the multiple peaks seen
in the observed radial distribution. We find that the smoothing method tends
to flatten and broaden the redshift distribution in our sample. A Gaussian
filter of smoothing length as large as $\sigma_s=450$\hmpc needs to be applied
to completely remove the spikes in the radial distribution. This, however
distorts the shape of radial distribution as shown in Fig. \ref{smoothing}.
An alternative way of estimating the expected $N(z)$ that we adopt here, is to
use the galaxy luminosity function. For volume-limited samples, one can
recover the radial distribution $N(z)$ as
\begin{equation}
N(z)dz=\int_{-\infty}^{M_{max}} \phi(M,z) \frac{dV_c}{dz} dM dz,
\end{equation}
where $\phi(M,z)$ is the galaxy luminosity function at redshift $z$ and $V_c$
is the comoving volume. In practice, to obtain $\phi(M,z)$, we first fitted
the evolution of the measured Schechter parameters in different redshift bins
by \citet{zucca09} with linear functions of $z$. We then use the functions
that best describe $\phi^*(z)$, $M^*(z)$, and $\alpha(z)$ to obtain
$\phi(M,z)$ at any redshift $z$, assuming a Schechter form of the luminosity
function. This method has the advantage of reducing sample variance effects in
the determination of the radial distribution by smoothing out the spurious
variations in the measured Schechter parameters from one redshift bin to
another. The resulting predicted radial distribution using this method is also
shown in Fig. \ref{smoothing}.

\subsection{Error estimation}
    
We estimate the errors in the correlation function measurements using the
blockwise bootstrap method \citep[e.g.][]{porciani02}, which provides a
reliable estimate of both the statistical and sample variance errors. We
verified their reliability by directly comparing with the ensemble average
scatter of a set of mock samples of the survey, and chose blockwise errors
because they produce a more stable covariance matrix. This internal-error
estimator consists in calculating the variance in the correlation functions
among a given number of realisations $N_{real}$ of the sample, consisting of a
random sequence of $N_{sub}$ equal-sized sub-volumes, allowing for
repetitions. To define the sub-volumes and because the transverse dimension of
the survey is small, we divided the sample into slices along the radial
direction and constrained the radial size of each slice to be larger than $5$
times $\pi_{max}$, as defined in Sect. \ref{method}. Therefore, for each
sample, we generated $N_{real}=800$ realisations by bootstrapping $N_{sub}=8$
slices of equal volume.

\citet{norberg09} studied the efficiency of different error estimators,
showing that this particular technique enables us to robustly recover the main
eigenvectors of the underlying covariance matrix and that by obtaining a large
number of realisations $N_{real}>3N_{sub}$, one can calculate variances that
agree with external estimators. However, their analysis is performed by
considering significantly larger volumes than the one probed by the zCOSMOS-Bright
sample, which prompted us to check directly the reliability of this method for
our specific survey using mock samples. We used 24 realistic mock samples of
the COSMOS survey provided by \citet{kitzbichler07} and based on the
Millennium dark matter N-body simulation \citep{springel05}. We applied to
them the detailed zCOSMOS-Bright observational strategy. For this purpose, we
used {\sc SSPOC} software \citep{bottini05} to mimic in the mock samples, the
selection of spectroscopic targets within the observed pointings. In addition,
we included the same redshift success rate as the real data \citep[see][for
  details]{iovino10}. We find that the blockwise bootstrap method, on average,
allows us to recover fairly well both the diagonal and off-diagonal terms of
the covariance matrix, although the diagonal errors are slightly
underestimated on large scales. This effect is however of the order of
$20-25\%$ on scales between $2$\hmpc and $10$\hmpc. Details of the error
analysis are presented in \citet{meneux09} and Porciani et al. (in
preparation).

\section{Results}

\begin{figure}
\includegraphics[width=84mm]{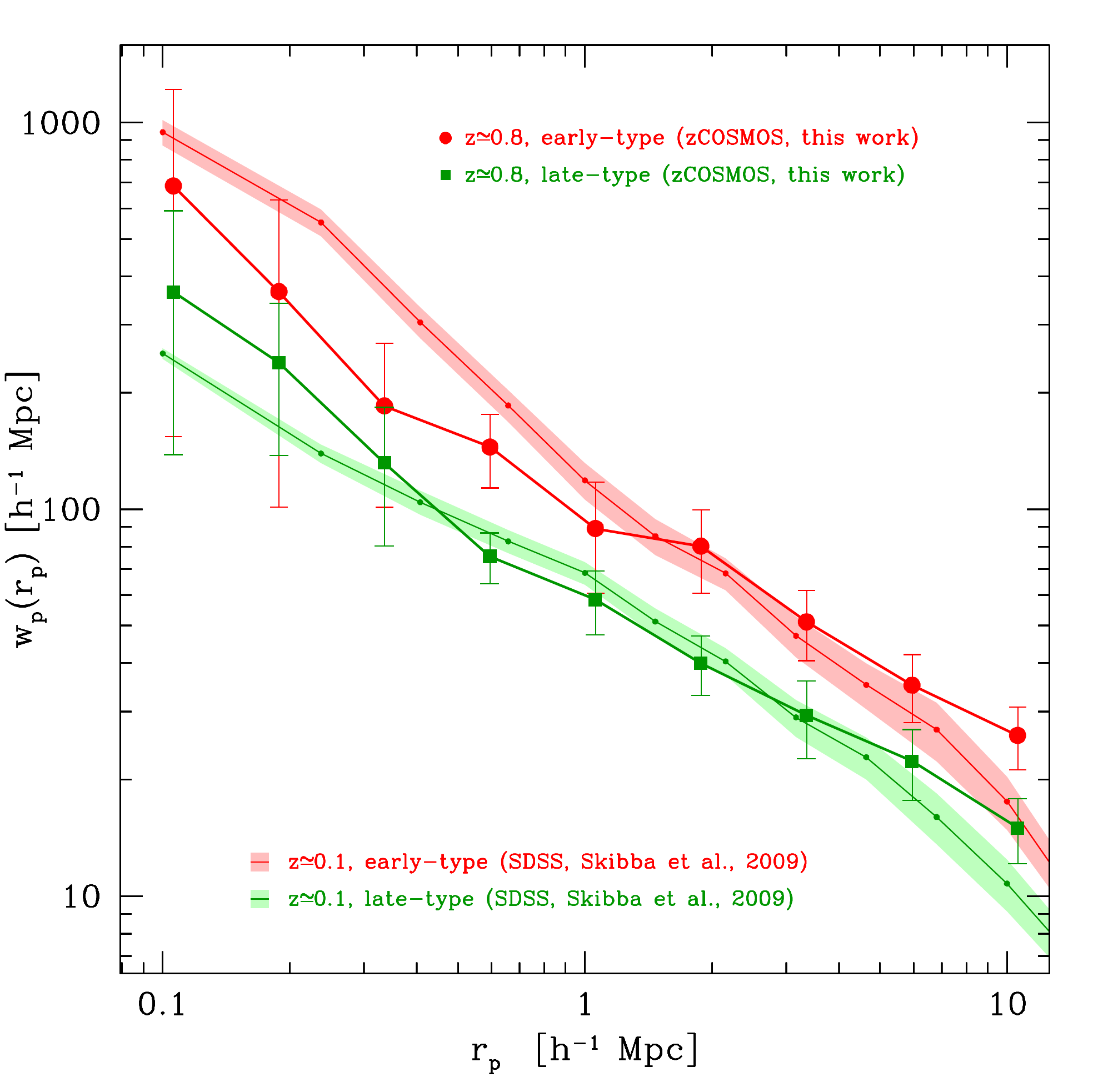}
\caption{The projected correlation functions of early-types (circles) and
  late-types (squares) in our volume-limited sample at $0.6<z<1.0$. The solid
  lines correspond to the measurements of \citet{skibba09} at $z\simeq0.1$.}
\label{morpho_res}
\end{figure}

We present in Fig. \ref{morpho_res} our measurement of the spatial clustering
of early- and late-type galaxies at ${\bar z}=0.81$, showing that at this
epoch, the former are already more strongly clustered than the latter.  The
observed segregation tends to only affect the amplitude of the projected
correlation function on all probed scales, corresponding to an almost
scale-independent relative clustering of early- to late-type galaxies that we
discuss in Sect.  \ref{sec:rbias}. We compare our measurements with those
obtained by \citet{skibba09} at $z\simeq0.1$ in the Sloan Digital Sky Survey
Data Release 6 \citep[SDSS,][]{adelman08} and based on visually classified
morphologies from the Galaxy Zoo sample \citep{lintott08}. We overplot in
Fig. \ref{morpho_res} the \wprp measurements they obtained for early and late
morphological types \citep[defined as $P_{el}>0.8$ and $P_{sp}>0.8$
  respectively, see][]{skibba09}, with absolute magnitudes of
$M_r<-20.5$. This magnitude cut selects galaxies with luminosity above $L^*$
at $z\simeq 0.1$ \citep{blanton03}, making these local measurements comparable
with ours. However, we cannot make a fully quantitative comparison as the
sample selection and morphological classifications are not exactly the
same. Comparison of the correlation function shapes shows some indications of
an increase with cosmic time in the relative difference in clustering between
early and late morphological classes for separations smaller than a few \hmpc,
although error bars are relatively large on these scales in our sample.

\begin{figure}
\includegraphics[width=84mm]{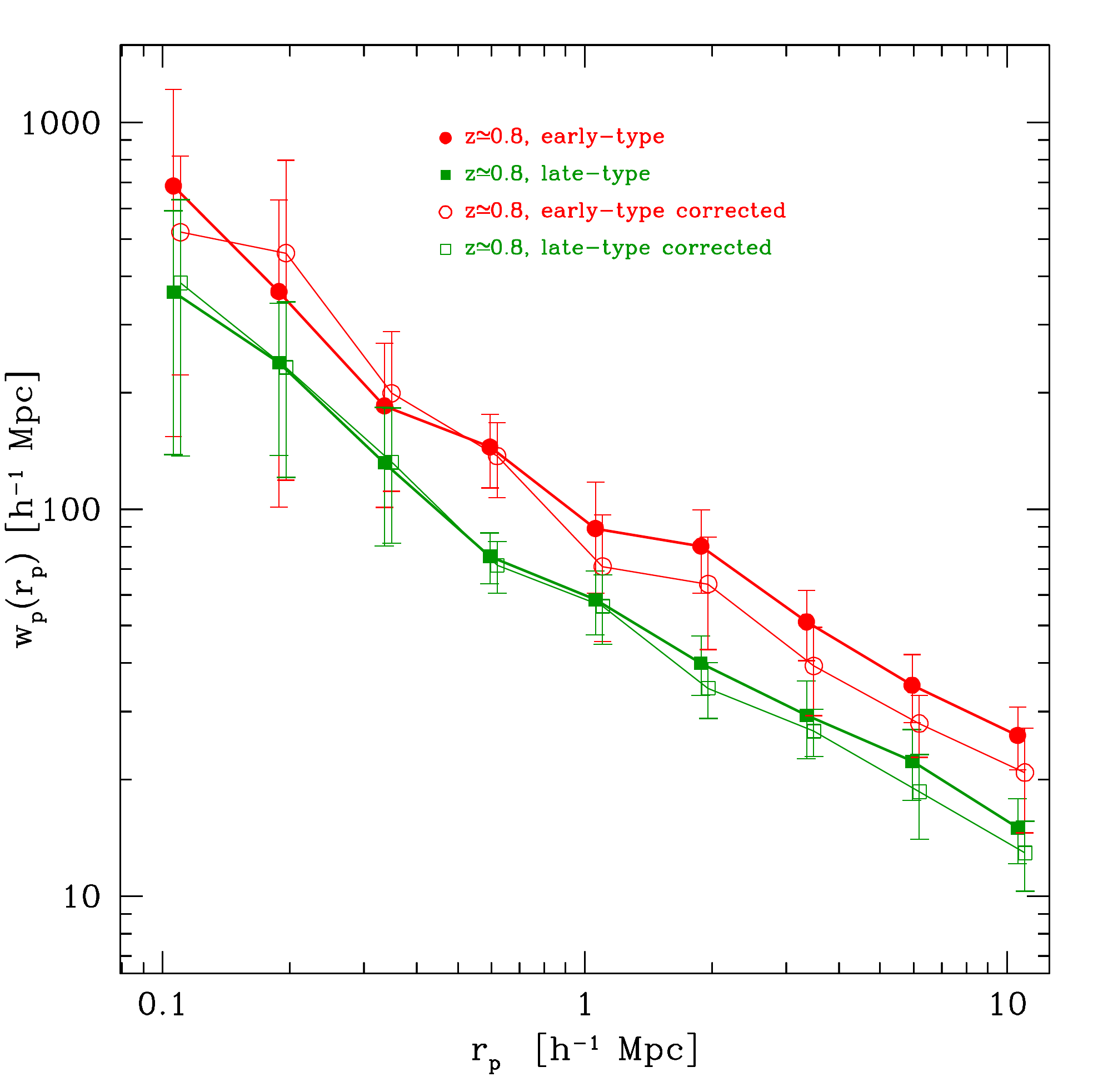}
\caption{The projected correlation functions of early-types (circles) and
  late-types (squares) in our volume-limited sample at $0.6<z<1.0$. The empty
  circles (for early types) and empty squares (for late types) are obtained
  while removing all the galaxies inhabiting the rich structures in excess in
  the zCOSMOS sample (see text for details). The latter points have been
  slightly displaced along the $r_p$-axis to improve the clarity of the
  figure.}
\label{morpho_struct}
\end{figure}

\subsection{Effect of the overabundance of high-density regions in the COSMOS field}

In the measurements of Fig. \ref{morpho_res}, the shape of the projected
correlation function at $r_p>1$\hmpc, for both early- and late-type galaxies,
is flatter than that for the SDSS, with approximately $w_p(r_p)\propto
r_p^{-1.6}$. Such behaviour was also noticed in previous analysis of the
zCOSMOS-Bright data, based on luminosity- and stellar mass-selected samples
\citep{meneux09} and has been recently explained as being due to an
overabundance of rich structures in the field, in particular at $z>0.6$
\citep{delatorre10}. This is quantified by an excess of galaxies in the
high-density tail of the overdensity probability distribution function (PDF)
of the sample, which is responsible for an enhancement of the clustering
signal on scales above $1-2$\hmpc. \citet{delatorre10} find that simply
removing galaxies inhabiting the 10\% densest environments brings the shape of
\wprp back into agreement with current $\Lambda CDM$ model predictions and
other datasets \citep{meneux08,meneux09}. We explore the significance of this
effect in our morphologically-selected samples performing exactly the same
operation here. The resulting \wprp are shown as empty symbols in
Fig. \ref{morpho_struct}.  This procedure reduces the amplitude of \wprp on
large scales, although the result is less dramatic than in the full sample
\citep{delatorre10}, and allows us to obtained better-behaved \wprp shapes on
these scales.

\subsection{Relative clustering of early- to late-type galaxies} \label{sec:rbias}

\begin{figure}
\includegraphics[width=84mm]{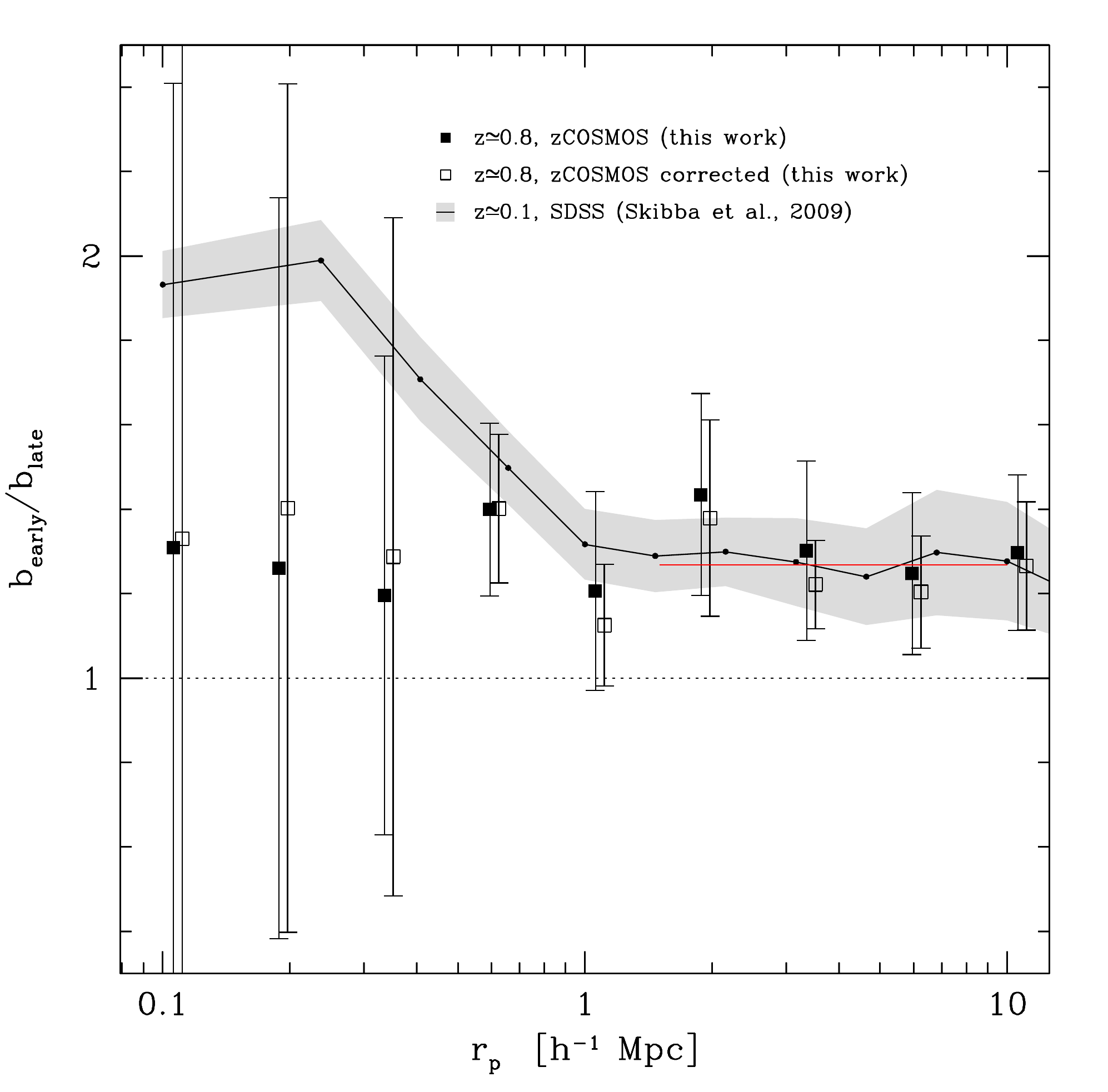}
\caption{Relative clustering of early- to late-type galaxies as a function of
  scale in our sample (symbols), compared to that at $z\simeq0.1$ obtained
  from \citet{skibba09} measurements (solid curve). The amplitude of the solid
  line defines the large-scale linear relative bias of early- to late-type
  galaxies, i.e., the plateau measured in the range $1.5~{\rm
    h^{-1}~Mpc}<r_p<10$\hmpc. The empty squares show the relative bias
  obtained while removing all the galaxies inhabiting the rich structures in
  excess in the zCOSMOS sample (see text for details).}
\label{brel}
\end{figure}

The observed relative clustering between different galaxy types can provide us
with valuable information on the properties of the galaxy formation bias
\citep[e.g.][]{narayanan00}. In general, local biasing models predict constant
scale-independent relative clustering on large scales
\citep{coles93,fry93,mann98,scherrer98,narayanan00}. This prediction remains
in any model where the galaxy type is correlated with the local density. We
note that here ``local'' refers to scales over which material in non-linear
structures has mixed during the cosmic history. This behaviour disagrees with
models of non-local bias in which, for instance, the galaxy formation
efficiency is coherently modulated over large scales because of ionising
radiation \citep{bower93} or suppressed in randomly distributed voids
\citep{babul91}. In these models, the large-scale relative clustering is
expected to be scale-dependent \citep{narayanan00}.

\begin{table*}
  \begin{minipage}{126mm}
  \caption{Large-scale relative bias of early- to late-type galaxies. These
    values are obtained by averaging the relative bias of early- to late-type
    galaxies between $1.5$\hmpc and $10$\hmpc.}
  \begin{tabular}{@{}ccccl}
  \hline
  ${\bar z}$ & selection & ${\bar L}/L^*$ & $b^{lin}_{rel}$ & reference\\
  \hline
  0.8 & $M_B<-20.4+(1.0-z)$ & $1.1$ & $1.27\pm0.16$ & this work \\
  0.1 & $M_r<-20.5$ & $1.1$ & $1.26\pm0.12$ & [1]~~\citet{skibba09} \\
  $\sim0$ & $M_B<-19.5$ & $1$ & $1.19\pm0.25$ & [2]~~\citet{willmer98} \\
  $\sim0$ & $M_{B}<-19.5$ & $1$ & $1.53\pm0.19$ & [3]~~\citet{guzzo97} \\
  $\sim0$ & $M_j\simeq-19.5$ & $0.9$ & $1.26\pm0.36$ & [4]~~\citet{loveday95} \\
  \hline
  \label{breltab}
  \end{tabular}

  \medskip
  For references [2], [3], and [4], the large-scale relative bias is
  calculated from the power-law best-fits to the \wprp of elliptical/S0
  (early-type) and spiral (late-type) galaxies; errors are obtained by
  propagating errors in the individual power-law parameters.
  \end{minipage}
\end{table*}

Figure \ref{brel} shows the relative clustering (or relative bias)
$b_{rel}(r_p)$ of early- to late-type galaxies in the range $0.1~{\rm
  h^{-1}~Mpc}<r_p<10$\hmpc obtained in our sample and from SDSS measurements
as
\begin{equation}
  b_{rel}=\left(\frac{w_p^{early}}{w_p^{late}}\right)^\frac{1}{2}=\left(\frac{w_p^{early}}{w_p^{m}}\frac{w_p^{m}}{w_p^{late}}\right)^\frac{1}{2}=\frac{b_{early}}{b_{late}},
\end{equation}
where $w_p^{early}$ and $b_{early}$ ($w_p^{late}$ and $b_{late}$) correspond
respectively to the projected correlation function and the bias of early-type
(late-type) galaxies with respect to the mass; $w_p^{m}$ is the projected
correlation function of mass. At $z\simeq0.1$, the relative bias shows a
scale-dependence in the regime where clustering is non-linear, i.e. on small
scales, but tends asymptotically to a constant value on larger scales. It
varies from $b_{rel}\simeq2$ at $r_p\simeq0.2$\hmpc to $b_{rel}\simeq1.3$ at
$r_p\simeq5$\hmpc, confirming previously reported clustering measurements in
the local Universe \citep{loveday95,hermit96,guzzo97,willmer98}. The observed
large-scale behaviour of the relative bias then supports a locally biased
galaxy formation scenario, in which the definition of early and late
morphological types may be directly or indirectly related to the local
environment. Moreover the large value of the relative bias on small scales,
i.e. below 1\hmpc, suggests that a fraction of early-type galaxies formed
recently and preferentially in dense environments. At $z\simeq0.8$, in
contrast, the scale dependence is much less important or absent. These results
may indicate that a fraction of early-type galaxies may have formed at epochs
later than $z\simeq0.8$ by the merging of late-type galaxies (perhaps through
major mergers), in relatively dense environments.

From the relative bias of early- to late-type galaxies, we derived the
large-scale linear relative bias $b^{lin}_{rel}$ by averaging $b_{rel}(r_p)$
on scales of $1.5~{\rm h^{-1}~Mpc}<r_p<10$\hmpc. Our measurement, corrected
for the effect of the excess of rich structures in the field, is reported in
Table \ref{breltab} where it is compared with $z\simeq0$
measurements. Interestingly, we find in our sample a value of $b^{lin}_{rel}$
similar to that measured in the local samples and in particular in the
SDSS. This finding suggests that most of the dependence of galaxy clustering
on morphology was already in place at $z\simeq0.8$ for $L>L^*$ galaxies.

\section{Summary and discussion}

We have measured the dependence of clustering on morphology at $0.6<z<1.0$ for
galaxies of luminosity greater than $L^*$.  For this purpose, we have used the
first epoch zCOSMOS-Bright spectroscopic sample of galaxies, our study
benefiting from the availability of high resolution HST imaging for the COSMOS
field. We have computed the projected correlations function in volume-limited
samples of two broad morphological classes, early types (elliptical/S0) and
late types (spiral/irregular) and have compared them to $z\simeq0.1$ SDSS
measurements. Our two main results can be summarised as follows:
\begin{enumerate}
\item We find that at $z\simeq0.8$, early-type galaxies exhibit a
  stronger clustering strength than late-type galaxies on scales from
  $0.1$\hmpc to $10$\hmpc.
\item Comparing our results to those for the SDSS for galaxies with comparable
  luminosities, shows that while the relative difference in clustering between
  early and late morphological classes seems to increase with cosmic time on
  scales smaller than a few \hmpc, the large-scale difference does not evolve
  significantly since $z\simeq0.8$. This indicates that a large fraction of
  early-type galaxies were already formed in intermediate and dense
  environments at this epoch.
\end{enumerate}

The observed difference in shape of the correlation function that we observe
for early- and late-type galaxies, can be interpreted within the framework of
halo occupation distribution models \citep[e.g.][]{cooray02}. In these models,
the galaxy correlation function is the sum of two contributions, one
dominating the small scales that characterises the clustering of galaxies
inside haloes (1-halo term), and a large-scale contribution, which
characterises the clustering of galaxies belonging to different haloes (2-halo
term). The prominence of the 1-halo term observed for early-type galaxies,
i.e. the enhancement of the observed correlation function on scales smaller
than or of the order of the typical halo radius ($1-2$\hmpc), implies that
these galaxies are on average hosted by massive haloes with larger virial
radii in dense environments \citep{abbas06}. The weaker but steeper 1-halo
term of late-type galaxies indicates that these galaxies instead may be hosted
on average by less massive haloes inhabiting low-density environments. This
trend is qualitatively consistent with $z\simeq0$ measurements of the
clustering of red and blue galaxies, that dominate respectively the early- and
late-type populations \citep[e.g.][]{zehavi05,skibba09c}.

Our work is complementary to those of \citet{tasca09} and \citet{kovac10} who
study within the same sample, the evolution of the fraction of early and late
morphological types with environment, as defined either by the continuous
overdensity field or group/field environments, respectively. The measurements
presented here are consistent with the picture emerging from these works. In
particular, \citet{tasca09} observed a flattening in the morphology-density
relation with increasing redshift at fixed luminosity, extending to morphology
similar evolutionary trends observed in the colour- and spectral-type-density
relations \citep{cucciati06,cooper06,grutzbauch10}. The finding that early-
and late-type galaxies tend to have increasingly similar probabilities of
populating high-density/group and low-density/field environments with
increasing redshift, is qualitatively consistent with the evolution in the
morphological dependence of clustering discussed here, where we find a
corresponding decrease in the relative small-scale clustering of early- to
late-type galaxies. The mass and environmental physical processes play an
important role in shaping the morphology-density relation and its evolution
with cosmic time \citep[e.g.][]{kovac10}. It is found that mass has a dominant
contribution at $z\simeq1$ in particular for relatively bright
luminosity-selected sample \citep{tasca09}. In contrast, in the local
Universe, one finds that only a part of this relation can be attributed to the
variation in the stellar-mass function with environment, with the dense
environment-related processes becoming more important \citep{bamford09}. Our
results corroborate this evolutionary picture. While the large-scale shape of
the relative bias has remained constant since $z\simeq1$, the small-scale
shape exhibits a significant evolution. This is due to an increase in the
number of small-scale early-type pairs with cosmic time, as a result of that
of the relative contribution of environmental physical processes in
transforming early- to late-type galaxies, especially in dense environments.

\section*{Acknowledgments}

We acknowledge the anonymous referee for his careful review of the paper and
helpful suggestions. Financial support from INAF and ASI through grants
PRIN-INAF-2007 and ASI/COFIS/WP3110 I/026/07/0 is gratefully acknowledged. JDS
is supported by World Premier International Research Center Initiative (WPI
Initiative), MEXT, Japan.

This work is based on observations undertaken at the European Southern
Observatory (ESO) Very Large Telescope (VLT) under Large Program 175.A-0839
and also on observations with the NASA/ESA Hubble Space Telescope, obtained at
the Space Telescope Science Institute, operated by the Association of
Universities for Research in Astronomy, Inc. (AURA), under NASA contract NAS
5Y26555, with the Subaru Telescope, operated by the National Astronomical
Observatory of Japan, with the telescopes of the National Optical Astronomy
Observatory, operated by the Association of Universities for Research in
Astronomy, Inc. (AURA), under cooperative agreement with the National Science
Foundation, and with the Canada-France- Hawaii Telescope, operated by the
National Research Council of Canada, the Centre National de la Recherche
Scientifique de France, and the University of Hawaii.

\bsp

\label{lastpage}


\begin{thebibliography}{}

\bibitem[\protect\citeauthoryear{Abbas 
\& Sheth}{2006}]{abbas06} Abbas U., Sheth R.~K., 2006, MNRAS, 372, 1749 

\bibitem[\protect\citeauthoryear{Abbas 
\& Sheth}{2007}]{abbas07} Abbas U., Sheth R.~K., 2007, MNRAS, 378, 641 

\bibitem[\protect\citeauthoryear{Abraham et 
al.}{1996}]{abraham96} Abraham R.~G., van den Bergh S., 
Glazebrook K., Ellis R.~S., Santiago B.~X., Surma P., Griffiths R.~E., 
1996, ApJS, 107, 1 

\bibitem[\protect\citeauthoryear{Abraham, van den Bergh, 
\& Nair}{2003}]{abraham03} Abraham R.~G., van den Bergh S., Nair P., 2003, ApJ, 588, 218 

\bibitem[\protect\citeauthoryear{Adelman-McCarthy et 
al.}{2008}]{adelman08} Adelman-McCarthy J.~K., et al., 2008, 
ApJS, 175, 297 

\bibitem[\protect\citeauthoryear{Babul 
\& White}{1991}]{babul91} Babul A., White S.~D.~M., 1991, MNRAS, 253, 31P 

\bibitem[\protect\citeauthoryear{Bamford et 
al.}{2009}]{bamford09} Bamford S.~P., et al., 2009, MNRAS, 393, 
1324

\bibitem[\protect\citeauthoryear{Benoist et 
al.}{1996}]{benoist96} Benoist C., Maurogordato S., da Costa 
L.~N., Cappi A., Schaeffer R., 1996, ApJ, 472, 452 

\bibitem[\protect\citeauthoryear{Blanton et 
al.}{2003}]{blanton03} Blanton M.~R., et al., 2003, ApJ, 592, 819 

\bibitem[\protect\citeauthoryear{Bottini et 
al.}{2005}]{bottini05} Bottini D., et al., 2005, PASP, 117, 996 

\bibitem[\protect\citeauthoryear{Bower et al.}{1993}]{bower93} 
Bower R.~G., Coles P., Frenk C.~S., White S.~D.~M., 1993, ApJ, 405, 403 

\bibitem[\protect\citeauthoryear{Brinchmann et 
al.}{1998}]{brinchmann98} Brinchmann J., et al., 1998, ApJ, 499, 112 

\bibitem[\protect\citeauthoryear{Capak et al.}{2007}]{capak07} 
Capak P., et al., 2007, ApJS, 172, 99 

\bibitem[\protect\citeauthoryear{Coil et al.}{2004}]{coil04} 
Coil A.~L., et al., 2004, ApJ, 609, 525 

\bibitem[\protect\citeauthoryear{Coil et al.}{2008}]{coil08} 
Coil A.~L., et al., 2008, ApJ, 672, 153

\bibitem[\protect\citeauthoryear{Coles}{1993}]{coles93} Coles 
P., 1993, MNRAS, 262, 1065 

\bibitem[\protect\citeauthoryear{Cooper et al.}{2006}]{cooper06} 
Cooper M.~C., et al., 2006, MNRAS, 370, 198 

\bibitem[\protect\citeauthoryear{Cooray 
\& Sheth}{2002}]{cooray02} Cooray A., Sheth R., 2002, PhR, 372, 1 

\bibitem[\protect\citeauthoryear{Cucciati et 
al.}{2006}]{cucciati06} Cucciati O., et al., 2006, A\&A, 458, 39 

\bibitem[\protect\citeauthoryear{Davis 
\& Geller}{1976}]{davis76} Davis M., Geller M.~J., 1976, ApJ, 208, 13 

\bibitem[\protect\citeauthoryear{Davis et al.}{1988}]{davis88} 
Davis M., Meiksin A., Strauss M.~A., da Costa L.~N., Yahil A., 1988, ApJ, 
333, L9 

\bibitem[\protect\citeauthoryear{de la Torre et 
al.}{2007}]{delatorre07} de la Torre S., et al., 2007, A\&A, 475, 443 

\bibitem[\protect\citeauthoryear{de la Torre et 
al.}{2010}]{delatorre10} de la Torre S., et al., 2010, MNRAS, 1322 

\bibitem[\protect\citeauthoryear{de Ravel et 
al.}{2009}]{deravel09} de Ravel L., et al., 2009, A\&A, 498, 379

\bibitem[\protect\citeauthoryear{Dressler}{1980}]{dressler80} 
Dressler A., 1980, ApJ, 236, 351 

\bibitem[\protect\citeauthoryear{Ellison et 
al.}{2010}]{ellison10} Ellison S.~L., Patton D.~R., Simard L., 
McConnachie A.~W., Baldry I.~K., Mendel J.~T., 2010, MNRAS, 407, 1514 

\bibitem[\protect\citeauthoryear{Fisher et al.}{1994}]{fisher94} 
Fisher K.~B., Davis M., Strauss M.~A., Yahil A., Huchra J.~P., 1994, MNRAS, 
267, 927 

\bibitem[\protect\citeauthoryear{Fry 
\& Gaztanaga}{1993}]{fry93} Fry J.~N., Gaztanaga E., 1993, ApJ, 413, 447 

\bibitem[\protect\citeauthoryear{Garilli et 
al.}{2008}]{garilli08} Garilli B., et al., 2008, A\&A, 486, 683 

\bibitem[\protect\citeauthoryear{Giovanelli, Haynes, 
\& Chincarini}{1986}]{giovanelli86} Giovanelli R., Haynes M.~P., Chincarini
  G.~L., 1986, ApJ, 300, 77 

\bibitem[\protect\citeauthoryear{Gr{\"u}tzbauch et 
al.}{2010}]{grutzbauch10} Gr{\"u}tzbauch R., Conselice C.~J., Varela 
J., Bundy K., Cooper M.~C., Skibba R., Willmer C.~N.~A., 2010, MNRAS, in press

\bibitem[\protect\citeauthoryear{Guzzo et al.}{1997}]{guzzo97} 
Guzzo L., Strauss M.~A., Fisher K.~B., Giovanelli R., Haynes M.~P., 1997, 
ApJ, 489, 37 

\bibitem[\protect\citeauthoryear{Guzzo et al.}{2000}]{guzzo00} 
Guzzo L., et al., 2000, A\&A, 355, 1 

\bibitem[\protect\citeauthoryear{Hamilton}{1988}]{hamilton88} 
Hamilton A.~J.~S., 1988, ApJ, 331, L59 

\bibitem[\protect\citeauthoryear{Hawkins et 
al.}{2003}]{hawkins03} Hawkins E., et al., 2003, MNRAS, 346, 78 

\bibitem[\protect\citeauthoryear{Hermit et al.}{1996}]{hermit96} 
Hermit S., Santiago B.~X., Lahav O., Strauss M.~A., Davis M., Dressler A., 
Huchra J.~P., 1996, MNRAS, 283, 709

\bibitem[\protect\citeauthoryear{Ilbert et al.}{2009}]{ilbert09} 
Ilbert O., et al., 2009, ApJ, 690, 1236 

\bibitem[\protect\citeauthoryear{Iovino et al.}{1993}]{iovino93} 
Iovino A., Giovanelli R., Haynes M., Chincarini G., Guzzo L., 1993, MNRAS, 
265, 21 

\bibitem[\protect\citeauthoryear{Iovino et 
al.}{2010}]{iovino10} Iovino A., et al., 2010, A\&A, 509, A40 

\bibitem[\protect\citeauthoryear{Kitzbichler 
\& White}{2007}]{kitzbichler07} Kitzbichler M.~G., White S.~D.~M., 2007, MNRAS, 376, 2 

\bibitem[\protect\citeauthoryear{Koekemoer et 
al.}{2007}]{koekemoer07} Koekemoer A.~M., et al., 2007, ApJS, 172, 
196 

\bibitem[\protect\citeauthoryear{Kova{\v c} et 
al.}{2010}]{kovac10} Kova{\v c} K., et al., 2010, ApJ, 718, 86 

\bibitem[\protect\citeauthoryear{Landy 
\& Szalay}{1993}]{landy93} Landy S.~D., Szalay A.~S., 1993, ApJ, 412, 64 

\bibitem[\protect\citeauthoryear{Le F{\`e}vre et 
al.}{2003}]{lefevre03} Le F{\`e}vre O., et al., 2003, SPIE, 4834, 173 

\bibitem[\protect\citeauthoryear{Le F{\`e}vre et 
al.}{2005}]{lefevre05} Le F{\`e}vre O., et al., 2005, A\&A, 439, 877 

\bibitem[\protect\citeauthoryear{Leauthaud et 
al.}{2007}]{leauthaud07} Leauthaud A., et al., 2007, ApJS, 172, 219 

\bibitem[\protect\citeauthoryear{Li et al.}{2006}]{li06} Li 
C., Kauffmann G., Jing Y.~P., White S.~D.~M., B{\"o}rner G., Cheng F.~Z., 
2006, MNRAS, 368, 21 

\bibitem[\protect\citeauthoryear{Lilly et al.}{1996}]{lilly96} 
Lilly S.~J., Le F{\`e}vre O., Hammer F., Crampton D., 1996, ApJ, 460, L1 

\bibitem[\protect\citeauthoryear{Lilly et al.}{2007}]{lilly07} 
Lilly S.~J., et al., 2007, ApJS, 172, 70 

\bibitem[\protect\citeauthoryear{Lilly et al.}{2009}]{lilly09} 
Lilly S.~J., et al., 2009, ApJS, 184, 218 

\bibitem[\protect\citeauthoryear{Lintott et 
al.}{2008}]{lintott08} Lintott C.~J., et al., 2008, MNRAS, 389, 
1179 

\bibitem[\protect\citeauthoryear{Lotz, Primack, 
\& Madau}{2004}]{lotz04} Lotz J.~M., Primack J., Madau P., 2004, AJ, 128, 163 

\bibitem[\protect\citeauthoryear{Loveday et 
al.}{1995}]{loveday95} Loveday J., Maddox S.~J., Efstathiou G., 
Peterson B.~A., 1995, ApJ, 442, 457 

\bibitem[\protect\citeauthoryear{Loveday, Tresse, 
\& Maddox}{1999}]{loveday99} Loveday J., Tresse L., Maddox S., 1999, MNRAS, 310, 281 

\bibitem[\protect\citeauthoryear{Madgwick et 
al.}{2003}]{madgwick03} Madgwick D.~S., et al., 2003, MNRAS, 344, 
847 

\bibitem[\protect\citeauthoryear{Mann, Peacock, 
\& Heavens}{1998}]{mann98} Mann R.~G., Peacock J.~A., Heavens A.~F., 1998, MNRAS, 293, 209 

\bibitem[\protect\citeauthoryear{McCracken et 
al.}{2008}]{mccracken08} McCracken H.~J., Ilbert O., Mellier Y., Bertin E., Guzzo L., Arnouts S., Le F{\`e}vre O., Zamorani G., 2008, A\&A, 479, 321 

\bibitem[\protect\citeauthoryear{Meneux et 
al.}{2006}]{meneux06} Meneux B., et al., 2006, A\&A, 452, 387 

\bibitem[\protect\citeauthoryear{Meneux et 
al.}{2008}]{meneux08} Meneux B., et al., 2008, A\&A, 478, 299 

\bibitem[\protect\citeauthoryear{Meneux et 
al.}{2009}]{meneux09} Meneux B., et al., 2009, A\&A, 505, 463 

\bibitem[\protect\citeauthoryear{Mo 
\& White}{1996}]{mo96} Mo H.~J., White S.~D.~M., 1996, MNRAS, 282, 347 

\bibitem[\protect\citeauthoryear{Narayanan, Berlind, 
\& Weinberg}{2000}]{narayanan00} Narayanan V.~K., Berlind A.~A., Weinberg D.~H., 2000, ApJ, 528, 1

\bibitem[\protect\citeauthoryear{Norberg et 
al.}{2001}]{norberg01} Norberg P., et al., 2001, MNRAS, 328, 64 

\bibitem[\protect\citeauthoryear{Norberg et 
al.}{2002}]{norberg02} Norberg P., et al., 2002, MNRAS, 332, 827 

\bibitem[\protect\citeauthoryear{Norberg et 
al.}{2009}]{norberg09} Norberg P., Baugh C.~M., Gazta{\~n}aga E., 
Croton D.~J., 2009, MNRAS, 396, 19 

\bibitem[\protect\citeauthoryear{Park et al.}{1994}]{park94} 
Park C., Vogeley M.~S., Geller M.~J., Huchra J.~P., 1994, ApJ, 431, 569 

\bibitem[\protect\citeauthoryear{Peebles}{1980}]{peebles80} 
Peebles P.~J.~E., 1980, The Large-Scale Structure of the Universe. Princeton University
Press, Princeton, NJ, p. 435

\bibitem[\protect\citeauthoryear{Pollo et 
al.}{2005}]{pollo05} Pollo A., et al., 2005, A\&A, 439, 887 

\bibitem[\protect\citeauthoryear{Pollo et 
al.}{2006}]{pollo06} Pollo A., et al., 2006, A\&A, 451, 409

\bibitem[\protect\citeauthoryear{Porciani 
\& Giavalisco}{2002}]{porciani02} Porciani C., Giavalisco M., 2002, ApJ, 565, 24 

\bibitem[\protect\citeauthoryear{Postman 
\& Geller}{1984}]{postman84} Postman M., Geller M.~J., 1984, ApJ, 281, 95

\bibitem[\protect\citeauthoryear{Scherrer 
\& Weinberg}{1998}]{scherrer98} Scherrer R.~J., Weinberg D.~H., 1998, ApJ, 504, 607 

\bibitem[\protect\citeauthoryear{Scoville et 
al.}{2007}]{scoville07} Scoville N., et al., 2007, ApJS, 172, 1

\bibitem[\protect\citeauthoryear{Sheth 
\& Tormen}{2002}]{sheth02} Sheth R.~K., Tormen G., 2002, MNRAS, 329, 61 

\bibitem[\protect\citeauthoryear{Skibba 
\& Sheth}{2009}]{skibba09c} Skibba R.~A., Sheth R.~K., 2009, MNRAS,
  392, 1080

\bibitem[\protect\citeauthoryear{Skibba et al.}{2006}]{skibba06} 
Skibba R., Sheth R.~K., Connolly A.~J., Scranton R., 2006, MNRAS, 369, 68 

\bibitem[\protect\citeauthoryear{Skibba et al.}{2009}]{skibba09} 
Skibba R.~A., et al., 2009, MNRAS, 399, 966 

\bibitem[\protect\citeauthoryear{Springel et 
al.}{2005}]{springel05} Springel V., et al., 2005, Natur, 435, 629 

\bibitem[\protect\citeauthoryear{Swanson et 
al.}{2008}]{swanson08} Swanson M.~E.~C., Tegmark M., Blanton M., 
Zehavi I., 2008, MNRAS, 385, 1635 

\bibitem[\protect\citeauthoryear{Tasca et 
al.}{2009}]{tasca09} Tasca L.~A.~M., et al., 2009, A\&A, 503, 379

\bibitem[\protect\citeauthoryear{Wang et al.}{2007}]{wang07} 
Wang L., Li C., Kauffmann G., De Lucia G., 2007, MNRAS, 377, 1419 

\bibitem[\protect\citeauthoryear{White et al.}{1987}]{white87} 
White S.~D.~M., Frenk C.~S., Davis M., Efstathiou G., 1987, ApJ, 313, 505 

\bibitem[\protect\citeauthoryear{White, Tully, 
\& Davis}{1988}]{white88} White S.~D.~M., Tully R.~B., Davis M., 1988, ApJ, 333, L45 

\bibitem[\protect\citeauthoryear{Willmer, da Costa, 
\& Pellegrini}{1998}]{willmer98} Willmer C.~N.~A., da Costa L.~N., Pellegrini P.~S., 1998, AJ, 115, 869 

\bibitem[\protect\citeauthoryear{Zehavi et al.}{2002}]{zehavi02} 
Zehavi I., et al., 2002, ApJ, 571, 172 

\bibitem[\protect\citeauthoryear{Zehavi et al.}{2005}]{zehavi05}
Zehavi I., et al., 2005, ApJ, 630, 1 

\bibitem[\protect\citeauthoryear{Zucca et 
al.}{2006}]{zucca06} Zucca E., et al., 2006, A\&A, 455, 879 

\bibitem[\protect\citeauthoryear{Zucca et 
al.}{2009}]{zucca09} Zucca E., et al., 2009, A\&A, 508, 1217 

\end{thebibliography}
\end{document}